\documentclass[preprint,aps,floats,nofootinbib]{revtex4}

\input epsf.tex
\def\DESepsf(#1 width #2){\epsfxsize=#2 \epsfbox{#1}}

\usepackage{graphicx}

\newcommand{\be}{\begin{equation}}
\newcommand{\ee}{\end{equation}}
\newcommand{\bea}{\begin{eqnarray}}
\newcommand{\eea}{\end{eqnarray}}

\def\thebibliography#1{\centerline{\bf REFERENCES}
  \list{[\arabic{enumi}]}{\settowidth\labelwidth{[#1]}\leftmargin
  \labelwidth\advance\leftmargin\labelsep\usecounter{enumi}}
\def\newblock{\hskip .11em plus .33em minus -.07em}\sloppy
  \clubpenalty4000\widowpenalty4000\sfcode`\.=1000\relax}

\begin{document}
\draft

\vspace*{0.5cm}

\title{A Critical Study of the $B \to K\pi$ Puzzle}

\author{ \vspace{0.5cm}
C.~S.~Kim$^{1,2}$\footnote{cskim@yonsei.ac.kr,  JSPS Fellow}, ~
Sechul~Oh$^3$\footnote{scoh@phya.yonsei.ac.kr}, ~
and ~
Chaehyun~Yu$^1$\footnote{chyu@cskim.yonsei.ac.kr}}

\affiliation{ \vspace{0.3cm}
$^1$Department of Physics, Yonsei University,
Seoul 120-479, Korea \\
$^2$Graduate School of Science, Hiroshima University, Higashi-Hiroshima, Japan
739-8536 \\
$^3$Natural Science Research Institute, Yonsei University,
Seoul 120-479, Korea
\vspace{1cm}}

\vspace*{0.5cm}

\begin{abstract}
\noindent In the light of new experimental results on $B \to K\pi$ decays, we critically study
the decay processes $B \to K \pi$ in a phenomenological way.
Using the quark diagram approach and the currently available data, we determine the allowed
values of the relevant theoretical parameters, corresponding to the electroweak (EW) penguin,
the color-suppressed tree contribution, etc.
In order to find the most likely values of the parameters in a statistically reliable way,
we use the $\chi^2$ minimization technique.
Our result shows that the current data for $B \to K\pi$ decays strongly indicate
(large) enhancements of both the EW penguin and the color-suppressed tree contributions.
In particular, the color-suppressed tree effect needs to be enhanced by about an order of
magnitude to fit the present data.
\end{abstract}
\maketitle

\section{Introduction}

{}From $B$ factory experiments such as Belle and BaBar, a tremendous amount of experimental
data on $B$ meson decays are being collected and provide new limits on previously known
observables with great precision as well as an opportunity to see very rare processes for
the first time.
Experimentally plenty of two-body hadronic $B$ decays have been observed and some of data
for these decay modes, such as $B \to K\pi$, are now quite precise, which leads to
a precision era for the study of two-body hadronic $B$ decays.

There are four different decay channels (and their anti-particle decay channels) for
$B \to K\pi$ processes, depending on the electric charge configuration:
$B^+ \to K^0 \pi^+$, $B^+ \to K^+ \pi^0$, $B^0 \to K^+ \pi^-$, and $B^0 \to K^0 \pi^0$.
All the $B \to K\pi$ modes have already been observed in experiment and their
CP-averaged branching ratios have been measured within a few percent errors
by the BaBar and Belle collaborations
\cite{HFAG,Bornheim:2003bv,Chao:2003ue,Aubert:2002jb,Aubert:2004dn,Aubert:2004km,Aubert:2004kn}.
The measurements of direct CP asymmetries for the $B \to K\pi$ modes had contained large
errors so that the results have not led to any decisive conclusions until recently
\cite{HFAG,Abe:LP05,Chen:2000hv,Aubert:2004qm,Chao:2004jy,Chao:2004mn,Abe:2004xp}.
But, the observations of the direct CP asymmetry in $B^0 \to K^{\pm} \pi^{\mp}$ have been
recently achieved at the 5.7$\sigma$ level by BaBar and Belle \cite{Aubert:2004qm,Chao:2004jy,Chao:2004mn}.
For the other $B \to K\pi$ modes, the experimental results of the direct CP asymmetries
still include large errors.
Certain experimental data ($e.g.$, the branching ratios (BRs)) for
$B \to K\pi$ are currently more precise than the theoretical model
predictions based on QCD factorization, perturbative  QCD (pQCD), and so on. Thus, these decay modes
can provide very useful information for improving the model calculations.
Therefore, at the same time, the model-independent study becomes
very important.

In the light of those new data, including the direct CP asymmetry in $B^0 \to K^{\pm} \pi^{\mp}$,
many works have been recently done to study
the implications of the data
\cite{Mishima:2004um,Buras:2004th,Charng:2004ed,He:2004ck,Wu:2004xx,Baek:2004rp,Carruthers:2004gj,
Nandi:2004dx,Morozumi:2004ea}.
The quark level subprocesses for $B \to K\pi$ decays are $b \to s q \bar q ~ (q = u,d)$ penguin
processes which are potentially sensitive to any new physics effects beyond the Standard
Model (SM).  Thus, with the currently available precision data, it is very important to investigate
these modes as generally and critically as possible.
In this work, we critically study the decay processes $B \to K \pi$ in a phenomenological
way.  In particular, by noticing that the current data for $B \to K \pi$ can be divided into two
groups (relatively precise ones and the other ones), to be conservative, we try to investigate
the implications of the current experimental results systematically in a few steps, as we shall
see later.
We are mainly interested in investigating whether the conventional SM predictions are consistent
with the current data.  Furthermore, if there are some deviations between the conventional
estimates and the experimental results, we intend to carefully identify the source of the deviations
and estimate how large the contribution from the source can be.  For this aim, we use
the topological amplitudes in the quark diagram approach and try to determine the allowed values
of the relevant theoretical parameters, corresponding to the electroweak (EW) penguin, the
color-suppressed tree contribution, and so on, by the current data.
We should emphasize that the parameter values determined in this way are model-independent.
Then, by comparing our result with the conventional SM predictions, we shall be able to verify
whether the current data indicate any new physics effects.
In order to find the most likely values of the theoretical parameters in a statically reliable way,
we will adopt the $\chi^2$ analysis.
In this work, we do not consider $B \to \pi \pi$ modes simultaneously  with $B \to K\pi$ modes, though
they can be connected to each other by using flavor SU(3) symmetry.
It is because we do not want that our analysis would be spoiled by the unknown effects of the flavor
SU(3) breaking.  Also, as it turns out, the data on $B \to K\pi$ provide enough information for
the analysis.

The paper is organized as follows.  The relevant formulas for $B \to K \pi$ modes are presented
in Sec. II.  In Sec. III, the experimental results for $B \to K \pi$ are summarized and their
implications are investigated.
In Sec. IV, the $\chi^2$ analysis using $B \to K \pi$ decays is presented.
We conclude the analysis in Sec. V.

\section{The relevant formulas for $B \to K \pi$ decay modes}

In order to specify our notation, let us first summarize the formulas for
the relevant decay amplitudes, BRs, direct and indirect (mixing-induced)
CP asymmetries.
The decay amplitudes for two-body hadronic $B$ decays can be represented
in terms of the basis of topological quark diagram contributions \cite{Gronau:1994rj}.
The relevant decay amplitudes for $B \to K\pi$ can be written as
\cite{Yoshikawa:2003hb}
\begin{eqnarray}
&& A^{0+} \equiv A(B^+ \to K^0 \pi^+)
 = V_{ub}^* V_{us} \tilde A +V_{tb}^* V_{ts} \tilde P , \\
&& A^{+0} \equiv A(B^+ \to K^+ \pi^0)
 = - {1 \over \sqrt{2}}
  \left[ V_{ub}^* V_{us} (\tilde T +\tilde C +\tilde A)
  +V_{tb}^* V_{ts} (\tilde P +\tilde P_{EW} +\tilde P_{EW}^C) \right], \\
&& A^{+-} \equiv A(B^0 \to K^+ \pi^-)
 = - \left[ V_{ub}^* V_{us} \tilde T
  +V_{tb}^* V_{ts} (\tilde P +\tilde P_{EW}^C) \right], \\
&& A^{00} \equiv A(B^0 \to K^0 \pi^0)
 = - {1 \over \sqrt{2}} \left[ V_{ub}^* V_{us} \tilde C
  -V_{tb}^* V_{ts} (\tilde P -\tilde P_{EW}) \right],
\label{decayamp}
\end{eqnarray}
where $V_{ij} ~(i=u, t; ~ j=s, b)$ are Cabibbo-Kobayashi-Maskawa (CKM) matrix elements
and the amplitudes $\tilde T$, $\tilde C$, $\tilde A$, $\tilde P$, $\tilde P_{EW}$,
and $\tilde P_{EW}^C$ are defined as
\begin{eqnarray}
&& \tilde T \equiv T +P_u +E_u -P_c -E_c ~,
\label{tildeT} \\
&& \tilde C \equiv C -P_u -E_u +P_c +E_c ~,
\label{tildeC} \\
&& \tilde A \equiv A +P_u +E_u -P_c -E_c ~,
\label{tildeA} \\
&& \tilde P \equiv P_t +E_t - P_c -E_c -{1 \over 3} P_{EW}^C
 +{2 \over 3} E_{EW}^C ~,
\label{tildeP} \\
&& \tilde P_{EW} \equiv P_{EW} +E_{EW}^C ~,
\label{tildePEW} \\
&& \tilde P_{EW}^C \equiv P_{EW}^C - E_{EW}^C ~.
\label{tildePEWC}
\end{eqnarray}
Here $T$ is a color-favored tree amplitude, $C$ is a color-suppressed tree,
$A$ is an annihilation, $P_i~(i = u,c,t)$ is a QCD penguin,
$E_i$ is a penguin exchange, $P_{EW}$ is a color-favored EW penguin,
$P_{EW}^C$ is a color-suppressed EW penguin, $E_{EW}^C$ is a color-suppressed
EW penguin exchange diagram.

Since the QCD penguin contribution is dominant in $B \to K\pi$ decays, the decay
amplitudes are rewritten as \cite{Yoshikawa:2003hb}
\begin{eqnarray}
&& A^{0+} = - |V_{tb}^* V_{ts}| \tilde P
 \left[ 1 -r_{_A} e^{i\delta^A} e^{i\phi_3} \right] ,
\label{A0p} \\
&& A^{+0} = {1 \over \sqrt{2}} |V_{tb}^* V_{ts}| \tilde P
 \left[ 1 -( r_{_T} e^{i\delta^T} +r_{_C} e^{i\delta^C} +r_{_A} e^{i\delta^A} ) e^{i\phi_3}
  +r_{_{EW}} e^{i\delta^{EW}} +r_{_{EW}}^C e^{i\delta^{EWC}} \right] ,
\label{Ap0} \\
&& A^{+-} = |V_{tb}^* V_{ts}| \tilde P
 \left[ 1 -r_{_T} e^{i\delta^T} e^{i\phi_3} +r_{_{EW}}^C e^{i\delta^{EWC}} \right] ,
\label{Apm} \\
&& A^{00} = - {1 \over \sqrt{2}} |V_{tb}^* V_{ts}| \tilde P
 \left[ 1 -r_{_{EW}} e^{i\delta^{EW}} +r_{_C} e^{i\delta^C} e^{i\phi_3} \right] ,
\label{A00}
\end{eqnarray}
where the ratios of each contribution to the dominant one are defined as
\begin{eqnarray}
&& r_{_A} = {|V_{ub}^* V_{us} \tilde A|  \over |V_{tb}^* V_{ts} \tilde P|} ~, ~~
   r_{_T} = {|V_{ub}^* V_{us} \tilde T|  \over |V_{tb}^* V_{ts} \tilde P|} ~, ~~
   r_{_C} = {|V_{ub}^* V_{us} \tilde C|  \over |V_{tb}^* V_{ts} \tilde P|} ~,  \\
&& r_{_{EW}} = {|\tilde P_{EW}| \over |\tilde P|} ~, ~~~
   r_{_{EW}}^C = {|\tilde P_{EW}^C| \over |\tilde P|} ~.
\end{eqnarray}
Here $\delta^X$ denotes the relative strong phase between each amplitude $\tilde X$
and the dominant $\tilde P$, and $\phi_3 ~ (\equiv \gamma)$ is the angle of the
unitarity triangle.
We note that there exists a conventional hierarchy among the above ratios:
\begin{equation}
1 > r_{_T} \sim r_{_{EW}} > r_{_C} \sim r_{_{EW}}^C > r_A ~.
\label{convhierarchy}
\end{equation}
For instance, in the pQCD approach, those ratios are roughly estimated as
\cite{Mishima:2004um,Keum:2000wi}
\begin{eqnarray}
&& r_{_T} \approx 0.21,~ r_{_{EW}} \approx 0.14,~ r_{_C} \approx 0.02,
~ r_{_{EW}}^C \approx 0.01, ~ r_A \approx 0.005 ~.
\label{convhierarchy2}
\end{eqnarray}
It is also known that within the SM, under flavor SU(3) symmetry,
the relation $\delta^T \approx \delta^{EW}$ holds to a good approximation
\cite{Neubert:1998jq}, which can be deduced from the fact that the topology of the
color-allowed tree diagram is similar to that of the EW penguin diagram.

Then the CP-averaged BRs are given by
\begin{eqnarray}
\bar {\cal B}^{0+} &\equiv& \bar {\cal B}(B^{\pm} \to K \pi^{\pm})
 \propto \frac{1}{2} \left[ |A^{0+}|^2 + |A^{0-}|^2 \right]  \nonumber \\
 &=& |V_{tb}^* V_{ts}|^2 |\tilde P|^2
  \left[ 1 -2 r_{_A} \cos{\delta^A} \cos{\phi_3} \right] ,
\label{B0p} \\
2 \bar {\cal B}^{+0} &\equiv& 2 \bar {\cal B}(B^{\pm} \to K^{\pm} \pi^0)
 \propto \left[ |A^{+0}|^2 + |A^{-0}|^2 \right] \nonumber \\
 &=& |V_{tb}^* V_{ts}|^2 |\tilde P|^2
  \left\{ 1 +2 r_{_{EW}} \cos{\delta^{EW}}
  +2 r_{_{EW}}^C \cos{\delta^{EWC}} \right.
 \nonumber \\
 &\mbox{}& \left. - 2 \left( r_{_T} \cos{\delta^T} +r_{_C} \cos{\delta^C}
   +r_{_A} \cos{\delta^A} \right) \cos{\phi_3} +r_{_T}^2 +r_{_{EW}}^2 +r_{_C}^2 \right.
\label{Bp0} \\
 &\mbox{}& \left. - 2 \left[ r_{_T} r_{_{EW}} \cos{(\delta^T -\delta^{EW})}
   +r_{_C} r_{_{EW}} \cos{(\delta^C -\delta^{EW})}
   -r_{_T} r_{_C} \cos{(\delta^T -\delta^C)} \right] \cos{\phi_3} \right\} ,
 \nonumber \\
\bar {\cal B}^{+-} &\equiv& \bar {\cal B}(B_d \to K^{\pm} \pi^{\mp})
 \propto \frac{1}{2} \left[ |A^{+-}|^2 + |A^{-+}|^2 \right]
 \nonumber \\
 &=& |V_{tb}^* V_{ts}|^2 |\tilde P|^2
  \left[ 1 -2 r_{_T} \cos{\delta^T} \cos{\phi_3}
  +2 r_{_{EW}}^C \cos{\delta^{EWC}} +r_{_T}^2  \right] ,
\label{Bpm} \\
2 \bar {\cal B}^{00} &\equiv& 2 \bar {\cal B}(B_d \to K \pi^0)
 \propto \left[ |A^{00}|^2 + |A^{00}|^2 \right]
 \nonumber \\
 &=& |V_{tb}^* V_{ts}|^2 |\tilde P|^2
  \left[ 1 -2 r_{_{EW}} \cos{\delta^{EW}} +2 r_{_C} \cos{\delta^C} \cos{\phi_3}
   +r_{_{EW}}^2 +r_{_C}^2 \right]
 \nonumber \\
 &\mbox{}& \left. -2 r_{_{EW}} r_{_C} \cos{(\delta^{EW} -\delta^C)} \cos{\phi_3}
  \right] .
\label{B00}
\end{eqnarray}
Here we neglect the $r^2$ terms which include tiny quantities $r_{_A}$ and
$r_{_{EW}}^C$.  However, because recent studies on two-body hadronic $B$ decays show
that the color-suppressed tree contribution could be enhanced to a large amount through
certain mechanisms \cite{Cheng:2004ru,Kim:2004hx,Keum:2003js}, we keep the $r^2$ terms
including $r_{_C}$, in order to take that possibility into account.
This treatment differs from that in Refs. \cite{Mishima:2004um,Yoshikawa:2003hb},
where all the $r^2$ terms including $r_{_A}$ and $r_{_{EW}}^C$ as well as $r_{_C}$ were
simply neglected.
In fact, we shall see that a large enhancement of the color-suppressed tree contribution
is indicated by the present experimental data for $B \to K\pi$ modes.

The ratios between the BRs for the $B \to K\pi $ modes can be also defined as
\begin{eqnarray}
&& R_1 = \frac{\tau^+ \bar{\cal B}^{+-}}{\tau^0 \bar{\cal B}^{0+}}
 = 1 - 2  r_{_T} \cos \delta^T \cos \phi_3 + r_{_T}^2 ,
\label{R1} \\
&& R_c = \frac{2\bar{\cal B}^{+0}}{\bar{\cal B}^{0+}}
 = 1 + 2 r_{_{EW}} \cos\delta^{EW}
 -2 \left( r_{_T} \cos\delta^T +r_{_C} \cos\delta^C \right) \cos\phi_3  \nonumber \\
&& \hspace{2.4cm} + r_{_T}^2 + r_{_{EW}}^2 +  r_{_C}^2
 +2 r_{_T} r_{_C} \cos (\delta^T - \delta^C)  \nonumber \\
&& \hspace{2.4cm} - 2 \left[ r_{_{EW}} r_{_T} \cos (\delta^{EW}-\delta^T)
 +r_{_{EW}} r_{_C} \cos (\delta^{EW}-\delta^C) \right] \cos \phi_3 ,
\label{Rc} \\
&& R_n = \frac{\bar{\cal B}^{+-}}{2\bar{\cal B}^{00}}
 = 1 + 2 r_{_{EW}} \cos \delta^{EW}
 -2 \left( r_{_T} \cos\delta^T +r_{_C} \cos\delta^C \right) \cos\phi_3 \nonumber \\
&& \hspace{2.4cm} + r_{_T}^2 - r_{_{EW}}^2 -  r_{_C}^2
 + 4 r_{_{EW}}^2 \cos^2 \delta^{EW}  \nonumber \\
&& \hspace{2.4cm} + 2 r_{_{EW}} \left[  r_{_C} \cos (\delta^{EW}-\delta^C)
 -2 r_{_T} \cos \delta^{EW}\cos\delta^T
 -4 r_{_C} \cos \delta^{EW}\cos \delta^C \right] \cos \phi_3\nonumber \\
&& \hspace{2.4cm} +4 r_{_C} \cos \delta^C
 \left( r_{_C} \cos\delta^C + r_{_T} \cos \delta^T \right) \cos^2 \phi_3 ,
\label{Rn}
\end{eqnarray}
where $\tau^+ ~(\tau^0)$ is a life time of $B^+ ~(B^0)$ and
$\tau^+ / \tau^0 = 1.086 \pm 0.017$ \cite{Eidelman:2004wy}.
We notice that $R_1$ depends only on $r_{_T}$, $\delta^T$, and $\phi_3$.
If $r_{_T}$ and $\phi_3$ can be determined by other observations, $R_1$
becomes dependent only on $\delta^T$ which is the relative strong phase
between the effective tree and the effective strong penguin contribution.
Subsequently, using the experimental value of $R_1$, one can determine
the value of $\delta^T$, as shown in the next section.

The direct CP asymmetries are given by
\begin{eqnarray}
{\cal A}_{CP}^{0+} &\equiv&
 \frac{{\cal B}(B^- \to \bar K^0 \pi^-) -{\cal B}(B^+ \to K^0 \pi^+)}
  {{\cal B}(B^- \to \bar K^0 \pi^-) +{\cal B}(B^+ \to K^0 \pi^+)}
 = -2 r_{_A} \sin{\delta^A} \sin{\phi_3},
\label{ACP0p}  \\
{\cal A}_{CP}^{+0} &\equiv&
 \frac{{\cal B}(B^- \to K^- \pi^0) -{\cal B}(B^+ \to K^+ \pi^0)}
  {{\cal B}(B^- \to K^- \pi^0) +{\cal B}(B^+ \to K^+ \pi^0)}  \nonumber \\
 &=& -2 \left[ r_{_T} \sin{\delta^T} -r_{_C} \sin{\delta^C}
   -r_{_A} \sin{\delta^A}  +r_{_T} r_{_{EW}} \sin{(\delta^T +\delta^{EW})}
   \right.  \nonumber \\
 &\mbox{}& \left. +r_{_C} r_{_{EW}} \sin{(\delta^C +\delta^{EW})} \right]
   \sin{\phi_3}  \nonumber \\
 &\mbox{}&  -\left[ 2 r_{_T} r_{_C} \sin{(\delta^T +\delta^C)}
   +r_{_T}^2 \sin{2\delta^T}
   +r_{_C}^2 \sin{2\delta^C} \right] \sin{2\phi_3},
\label{ACPp0}  \\
{\cal A}_{CP}^{+-} &\equiv&
 \frac{{\cal B}(\bar B^0 \to K^- \pi^+) -{\cal B}(B^0 \to K^+ \pi^-)}
  {{\cal B}(\bar B^0 \to K^- \pi^+) +{\cal B}(B^0 \to K^+ \pi^-)}
 \nonumber \\
 &=& -2 r_{_T} \sin{\delta^T} \sin{\phi_3} -r_{_T}^2 \sin{2\delta^T} \sin{2\phi_3},
\label{ACPpm} \\
{\cal A}_{CP}^{00} &\equiv&
 \frac{{\cal B}(\bar B^0 \to \bar K^0 \pi^0) -{\cal B}(B^0 \to K^0 \pi^0)}
  {{\cal B}(\bar B^0 \to \bar K^0 \pi^0) +{\cal B}(B^0 \to K^0 \pi^0)}
 \nonumber \\
 &=& 2 \left[ r_{_C} \sin{\delta^C}
  +r_{_{EW}} r_{_C} \sin{(\delta^{EW} +\delta^C)} \right] \sin{\phi_3}
  -r_{_C}^2 \sin{2\delta^C} \sin{2\phi_3}.
\label{ACP00}
\end{eqnarray}
Notice that considering the conventional hierarchy given in (\ref{convhierarchy})
and (\ref{convhierarchy2}), the direct CP asymmetries ${\cal A}_{CP}^{+0}$ (\ref{ACPp0})
and ${\cal A}_{CP}^{+-}$ (\ref{ACPpm}) are expected to be almost same including their
{\it signs}, because the dominant contribution to them is identical.
However, the current experimental data show that ${\cal A}_{CP}^{+0}$ and
${\cal A}_{CP}^{+-}$ are quite different from each other and even have the opposite
signs to each other, as shown in Table~\ref{table:1}.

The time-dependent CP asymmetry for $B^0 \to K_{_S} \pi^0$ is defined as
\begin{eqnarray}
{\cal A}_{K_{_S} \pi^0} (t)
 &\equiv& \frac{\Gamma(\bar B^0 (t) \to K_{_S} \pi^0) -\Gamma(B^0 (t) \to K_{_S} \pi^0)}
  {\Gamma(\bar B^0 (t) \to K_{_S} \pi^0) +\Gamma(B^0 (t) \to K_{_S} \pi^0)}  \nonumber \\
 &\equiv& S_{K_{_S} \pi^0} \sin(\Delta m_d ~t) -C_{K_{_S} \pi^0} \cos(\Delta m_d ~t) ~,
\end{eqnarray}
where $\Gamma$ denotes the relevant decay rate and $\Delta m_d$ is the mass
difference between the two $B^0$ mass eigenstates.
The $S_{K_{_S} \pi^0}$ and $C_{K_{_S} \pi^0}$ are CP violating parameters.
In the case that the tree contributions are neglected for $B^0 \to K_{_S} \pi^0$,
the mixing-induced CP violating parameter $S_{K_{_S} \pi^0}$ is equal to $\sin(2\phi_1)$
[$\phi_1 ~ (\equiv \beta)$ is the angle of the unitarity triangle].
The expression for $S_{K_{_S} \pi^0}$ (up to $r^2$ order) is given by
\begin{eqnarray}
S_{K_{_S} \pi^0} &=& \sin (2\phi_1) \left( 1 -2 r_{_C}^2 \sin^2 \phi_3 \right)
 +\cos (2\phi_1) \left[ 2 r_{_C} \cos \delta^C \sin \phi_3  \right. \nonumber \\
&& \left. +2 r_{_{EW}} r_{_C} \cos (\delta^{EW} +\delta^C ) \sin \phi_3
   -r_{_C}^2 \cos (2\delta^C) \sin (2\phi_3) \right].
\label{Skpi}
\end{eqnarray}
The measured value of $S_{K_{_S} \pi^0}$ (Table~\ref{table:1}) is different from the
well-established value of $\sin(2\phi_1) =0.725 \pm 0.037$ measured through
$B \to J/\psi K^{(*)}$ \cite{HFAG}.
It may indicate that the subleading terms including $r_{_C}$ and $r_{_{EW}}$ in Eq. (\ref{Skpi})
play an important role.

\section{The $B \to K \pi$ puzzle and its implication}

We first summarize the present status of the experimental results on $B \to K\pi$
modes in Table I, which includes the BRs, the direct CP asymmetries
$({\cal A}_{CP})$, and the mixing-induced CP asymmetry $(S_{K_s \pi^0})$.
We see that the averages of the current experimental values for the BRs include
only a few percent errors.
Furthermore, the direct CP asymmetry in $B^0 \to K^{\pm} \pi^{\mp}$ has been recently
observed by the BaBar and Belle collaborations whose values
are in good agreement with each other (Table I): the world average value is
\begin{equation}
{\cal A}_{CP}^{+-} = -0.115 \pm 0.018 ~.
\end{equation}
The direct CP asymmetry data for the other $B \to K\pi$ modes involve large
uncertainties.
We also present the values of $R_1$, $R_c$, and $R_n$, defined in Eqs. (\ref{R1}) $-$
(\ref{Rn}), which are obtained from the experimental results given in Table I:
\begin{eqnarray}
&& R_1 = 0.82 \pm 0.06 ~,
\label{R1:data} \\
&& R_c = 1.00 \pm 0.09 ~,
\label{Rc:data} \\
&& R_n = 0.82 \pm 0.08 ~.
\label{Rn:data}
\end{eqnarray}

It has been also claimed that within the SM, $R_c - R_n \approx 0$ \cite{Buras:2004th,Buras:2003dj}.
{}From Eqs. (\ref{Rc}) and (\ref{Rn}), it is indeed clear that $R_c \approx R_n$, if the $r^2$
order terms including $r_{_{EW}}$ or $r_{_C}$ are negligible.
In other words, any difference between $R_c$ and $R_n$ would arise from the contributions
from the subdominant $r^2$ order terms including $r_{_{EW}}$ or $r_{_C}$.
The above experimental data show the pattern $R_c > R_n$ \cite{Buras:2004th,Buras:2003dj},
which would imply the enhancement of the EW penguin and/or the color-suppressed tree
contributions.  We will investigate the implication of the data below.

We remind that assuming the conventional hierarchy as in Eqs. (\ref{convhierarchy})
and (\ref{convhierarchy2}), ${\cal A}_{CP}^{+0}$ is expected to be almost
the same as ${\cal A}_{CP}^{+-}$: in particular, they would have the {\it same} sign.
However, the data show that ${\cal A}_{CP}^{+0}$ differs by 3.5$\sigma$ from
${\cal A}_{CP}^{+-}$.
This is a very interesting observation with the new measurements of
${\cal A}_{CP}^{+-}$ by BaBar and Belle, even though the measurements of
${\cal A}_{CP}^{+0}$ still include sizable errors.
One may need to explain on the theoretical basis how this feature can happen.

\begin{table}
\caption{Experimental data on the CP-averaged branching ratios ($\bar {\cal B}$
in units of $10^{-6}$), the direct CP asymmetries (${\cal A}_{CP}$), and
the mixing-induced CP asymmetry ($S_{K_s \pi^0}$) for $B \to K\pi$ modes.
The $S_{K_s \pi^0}$ is equal to $\sin(2\phi_1)$ in the case that tree
amplitudes are neglected for $B^0 \to K_s \pi^0$
\cite{HFAG,Bornheim:2003bv,Chao:2003ue,Aubert:2002jb,Aubert:2004dn,Aubert:2004km,
Aubert:2004kn,Abe:LP05,Chen:2000hv,Aubert:2004qm,Chao:2004jy,Chao:2004mn,Abe:2004xp}.}
\smallskip
\begin{tabular}{|c|c|c|c|c|}
\hline
  & CLEO & Belle & BaBar & Average  \\
\hline
$\bar {\cal B}(B^{\pm} \to K^0 \pi^{\pm})$ & $18.8^{+3.7 +2.1}_{-3.3 -1.8}$
 & $22.0 \pm 1.9 \pm 1.1$ & $26.0 \pm 1.3 \pm 1.0$ & $24.1 \pm 1.3$  \\
$\bar {\cal B}(B^{\pm} \to K^{\pm} \pi^0)$ & $12.9^{+2.4 +1.2}_{-2.2 -1.1}$
 & $12.0 \pm 1.3^{+1.3}_{-0.9}$ & $12.0 \pm 0.7 \pm 0.6$ & $12.1 \pm 0.8$  \\
$\bar {\cal B}(B^0 \to K^{\pm} \pi^{\mp})$ & $18.0^{+2.3 +1.2}_{-2.1 -0.9}$
 & $18.5 \pm 1.0 \pm 0.7$ & $19.2 \pm 0.6 \pm 0.6$ & $18.9 \pm 0.7$  \\
$\bar {\cal B}(B^0 \to K^0 \pi^0)$ & $12.8^{+4.0 +1.7}_{-3.3 -1.4}$
 & $11.7 \pm 2.3^{+1.2}_{-1.3}$ & $11.4 \pm 0.9 \pm 0.6$ & $11.5 \pm 1.0$  \\
\hline
${\cal A}_{CP}^{0+}$ & $+0.18 \pm 0.24 \pm 0.02$ & $+0.05 \pm 0.05 \pm 0.01$
 & $-0.09 \pm 0.05 \pm 0.01$ & $-0.02 \pm 0.04$   \\
${\cal A}_{CP}^{+0}$ & $-0.29 \pm 0.23 \pm 0.02$ & $+0.04 \pm 0.04 \pm 0.02$
 & $+0.06 \pm 0.06 \pm 0.01$ & $+0.04 \pm 0.04$   \\
${\cal A}_{CP}^{+-}$ & $-0.04 \pm 0.16 \pm 0.02$ & $-0.113 \pm 0.022 \pm 0.008$
 & $-0.133 \pm 0.030 \pm 0.009$ & $-0.115 \pm 0.018$\footnote{This average also
 includes the CDF result: $-0.04 \pm 0.08 \pm 0.01$.}   \\
${\cal A}_{CP}^{00}$ & $-$ & $+0.16 \pm 0.29 \pm 0.05$
 & $-0.06 \pm 0.18 \pm 0.03$ & $+0.001 \pm 0.155$   \\
\hline
$S_{K_s \pi^0}$ & $-$ & $+0.30 \pm 0.59 \pm 0.11$
 & $+0.35^{+0.30}_{-0.33} \pm 0.04$ & $+0.34 \pm 0.29$  \\
\hline
\end{tabular}
\label{table:1}
\end{table}

Based on the current experimental data shown in Table I, we critically
investigate their implications to the underlying theory on the $B \to K\pi$
processes.
There are nine observables available for the $B \to K\pi$ modes as shown
in Table~\ref{table:1}, but if the expectedly very small annihilation term $r_{_A}$
is neglected, the observable ${\cal A}_{CP}^{0+}$ becomes irrelevant and only
eight observables remain relevant.
There are also eight parameters ($|P|$, $r_{_T}$, $r_{_{EW}}$, $r_{_C}$,
$\delta^T$, $\delta^{EW}$, $\delta^C$, $\phi_3$) relevant to the above
observables, neglecting the very small terms including $r_{_A}$ and $r_{_{EW}}^C$
[see Eqs.~(\ref{B0p})~$-$~(\ref{ACP00})].
In our numerical analysis, we take into account the above eight parameters.
[Equivalently, we can use only seven observables ($i.e.$, three $R_i ~(i =1, c, n)$
and the CP asymmetries in Table~\ref{table:1}),
and take into account the seven parameters (except for $|P|$ among
the above eight parameters).]

We remind that among the data shown in Table I, five of them, such as the BRs
and ${\cal A}_{CP}^{+-}$, involve relatively small uncertainties, but the others
still include large errors.
Based on this observation, we consider four different cases as follows.
{\bf (i)} We first use only the four BRs in our analysis.
{\bf (ii)} Then, the data for ${\cal A}_{CP}^{+-}$ are also considered in addition to the
BRs.  Since the observables used in the cases {\bf (i)} and {\bf (ii)} are measured
relatively accurately, the result from these cases would provide more solid
implications.
{\bf (iii)} Then, we also use the data for $S_{K_s \pi^0}$ besides those used in the case
{\bf (ii)}.
{\bf (iv)} Finally we use all the currently available data.

\begin{figure}
 \centerline{ \DESepsf(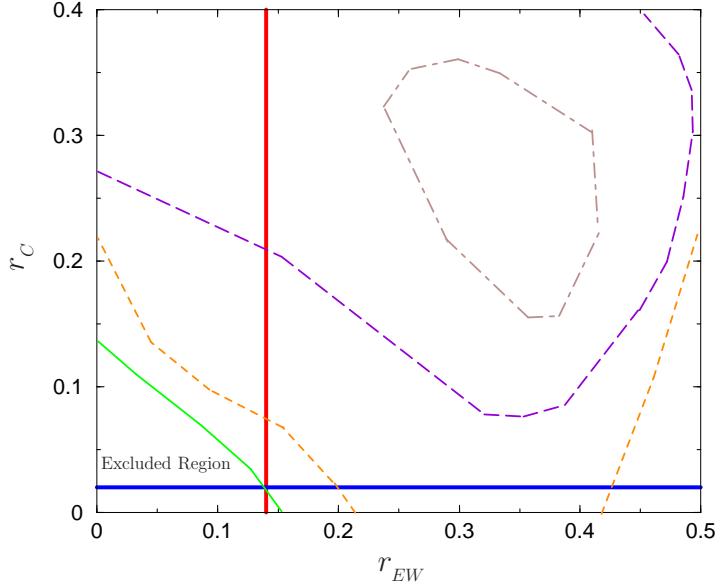 width 9cm)}
 \caption{The excluded region for $r_{_C}$ and $r_{_{EW}}$.  The bold straight
 (parallel and vertical) lines denote the conventional values of
 $r_{_C} \approx 0.02$ and $r_{_{EW}} \approx 0.14$, respectively.
 The solid, dotted, dashed, dot-dashed lines, respectively, correspond to the
 case of using only 4 BRs, $4~{\rm BRs} + {\cal A}_{CP}^{+-}$,
 $4~{\rm BRs} + {\cal A}_{CP}^{+-} + S_{K_s \pi^0}$, and all the available
 data.  Here $54^\circ \leq \phi_3 \leq 67^\circ$ and $0 \leq r_{_T} \leq 0.4$
 were used.}
\label{fig:1}
\end{figure}

In Fig.~\ref{fig:1}, we show the excluded region for $r_{_C}$ and $r_{_{EW}}$ by the
current data for the cases {\bf (i)}$-${\bf (iv)}.
The graph has been obtained by using the current data given in Table I as constraints
and directly solving Eqs.~(\ref{B0p})$-$(\ref{B00}), (\ref{ACPp0})$-$(\ref{ACP00}),
and (\ref{Skpi}) for $54^\circ \leq \phi_3 \leq 67^\circ$ and $0 \leq r_{_T} \leq 0.4$.
[Here we use the value of $\phi_3$ given by the unitarity triangle fit \cite{Bona:2005vz}.
In order to study the effect of $r_{_T}$ to the result, we vary the value of
$r_{_T}$ from 0 to 0.4.]
The bold straight (parallel and vertical) lines denote the conventional values of
$r_{_C} \approx 0.02$ and $r_{_{EW}} \approx 0.14$, respectively.
The solid, dotted, dashed, dot-dashed lines, respectively, correspond to each
case of {\bf (i)}$-${\bf (iv)} in order: $i.e.$, each case of using {\bf (i)} only 4 BRs,
{\bf (ii)} $4~{\rm BRs} + {\cal A}_{CP}^{+-}$,
{\bf (iii)} $4~{\rm BRs} + {\cal A}_{CP}^{+-} + S_{K_s \pi^0}$,
and {\bf (iv)} all the available data.

One should keep in mind that in the cases {\bf (i)}$-${\bf (iii)} only the ``excluded'' regions
for $r_{_C}$ and $r_{_{EW}}$ are meaningful {\it at 1$\sigma$ level} in Fig.~\ref{fig:1},
because in these cases the number of parameters are larger than that of the relevant
equations so that the other regions (except for the excluded regions) do not exactly
mean the ``allowed'' regions for $r_{_C}$ and $r_{_{EW}}$ at 1$\sigma$ level.

The result shows that when the data only for the BRs of $B \to K \pi$ modes are taken
into account [solid line; case {\bf (i)}], the conventional SM predictions of both
$r_{_{EW}} \approx 0.14$ and $r_{_C} \approx 0.02$ are not completely excluded by the data.
But, we should add a remark that this fact holds only for $r_{_T} \approx 0.4$ (which is larger
than the conventional SM estimate of $r_{_T} \approx 0.2$), because for smaller $r_{_T}$
larger values of $r_{_{EW}}$ and $r_{_C}$ are excluded: $e.g.$, for $r_{_T} =0.3$ or smaller,
$r_{_{EW}} \approx 0.14$ together with $r_{_C} \approx 0.02$ is no more allowed.
In other words, for the conventional value of $r_{_T} \approx 0.2$, the data for the BRs
indicates that the EW penguin and/or the color-suppressed tree term need(s) to be enhanced.
[This conclusion equivalently holds when the data only for $R_i ~(i = 1,c,n)$ given in
Eqs. (\ref{R1:data})~$-$~(\ref{Rn:data}) are considered.]

When the constraint from the measured $A_{CP}^{+-}$ is added to the case {\bf (i)}, the
conventional values of $r_{_{EW}}$ and $r_{_C}$ are not allowed at the same time (dotted line).
For instance, if $r_{_{EW}} \approx 0.14$, then the values of $r_{_C}$ smaller than 0.07
are excluded.  On the other hand, if $r_{_C} \approx 0.02$, then the values of $r_{_{EW}}$
smaller than 0.19 are not allowed by the data.
It would be also possible that both $r_{_{EW}}$ and $r_{_C}$ are simultaneously enhanced:
$e.g.$, $r_{_{EW}} \approx 0.17$ and $r_{_C} \approx 0.05$ are not excluded in this case {\bf (ii)}.
Thus, we see that even in this conservative case of considering only 5 (relatively precisely
measured) observables, the data strongly indicate a sizable enhancement of the EW
penguin term $r_{_{EW}}$ or the color-suppressed tree term $r_{_C}$, or both of them.

The dashed line shows the result for the case of adding one more constraint from
$S_{K_s \pi^0}$ to the case {\bf (ii)}.
It is interesting to note that the values of $r_{_C}$ smaller than 0.08 are completely
excluded in this case {\bf (iii)}, independent of values of $r_{_{EW}}$.

Finally we consider all the available data for $B \to K\pi$ modes shown in Table I.
In this case, the number of parameters are the same as that of the relevant equations so
that one can determine the ``allowed'' values of the parameters by solving the equations
numerically.  In order to find the allowed region for the parameters,
when combining all the data, we carefully regulate the errors in the data to be within
1$\sigma$ in total.
The result is represented as the dot-dashed line [case {\bf (iv)}].  The allowed values for
$r_{_{EW}}$ and $r_{_C}$ are limited to a rather small region at 1$\sigma$ level: roughly,
$0.24 \leq r_{_{EW}} \leq 0.41$ and $0.16 \leq r_{_C} \leq 0.36$.
Notice that {\it simultaneous} enhancements of $r_{_C}$ and $r_{_{EW}}$ are indicated in this
case.
In order to confirm this result in a statistically more reliable way, we shall use the
$\chi^2$ minimization technique in the next section.

\begin{figure}[tb]
   \includegraphics[scale=0.45]{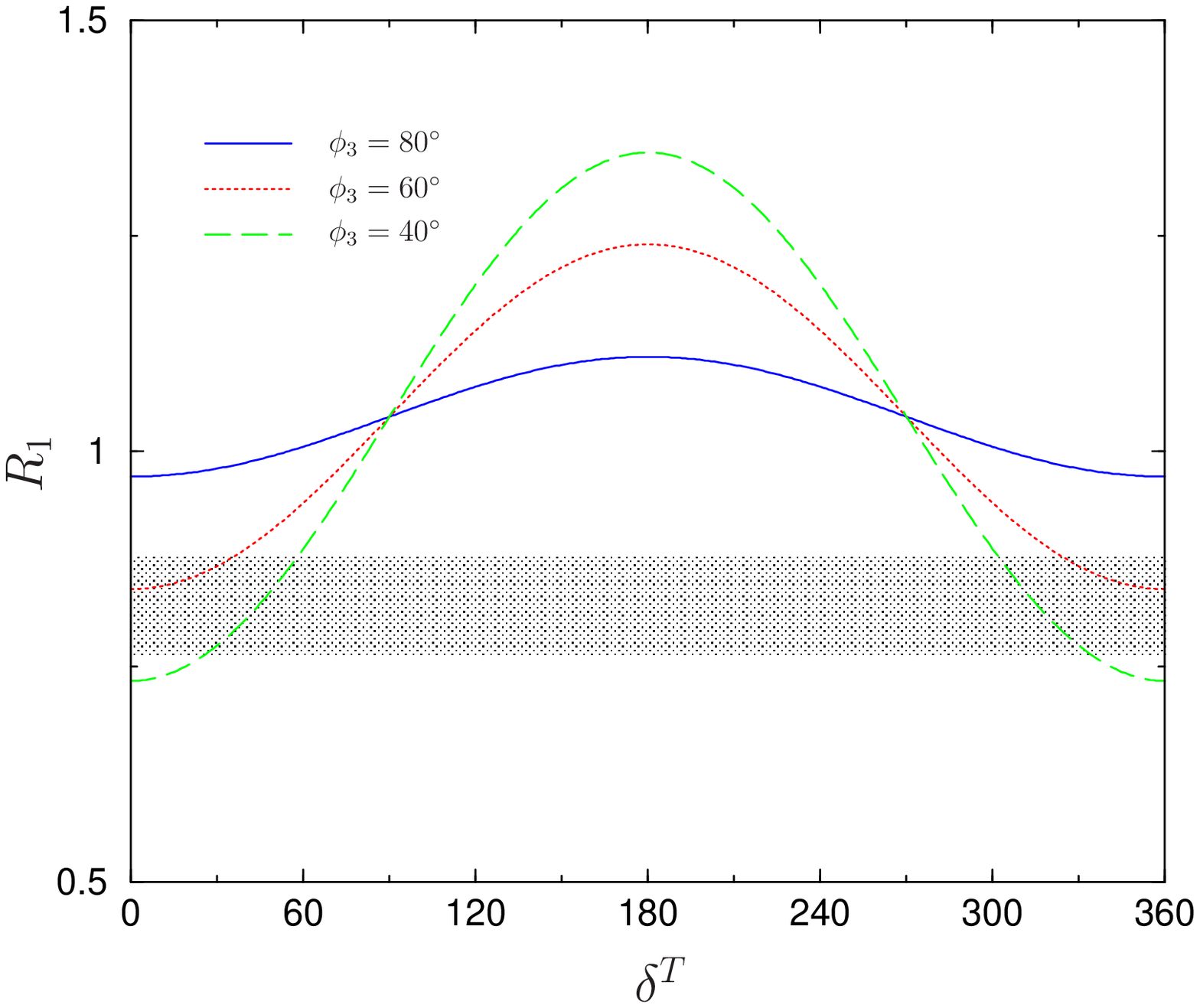}
 \hspace{1cm}
   \includegraphics[scale=0.45]{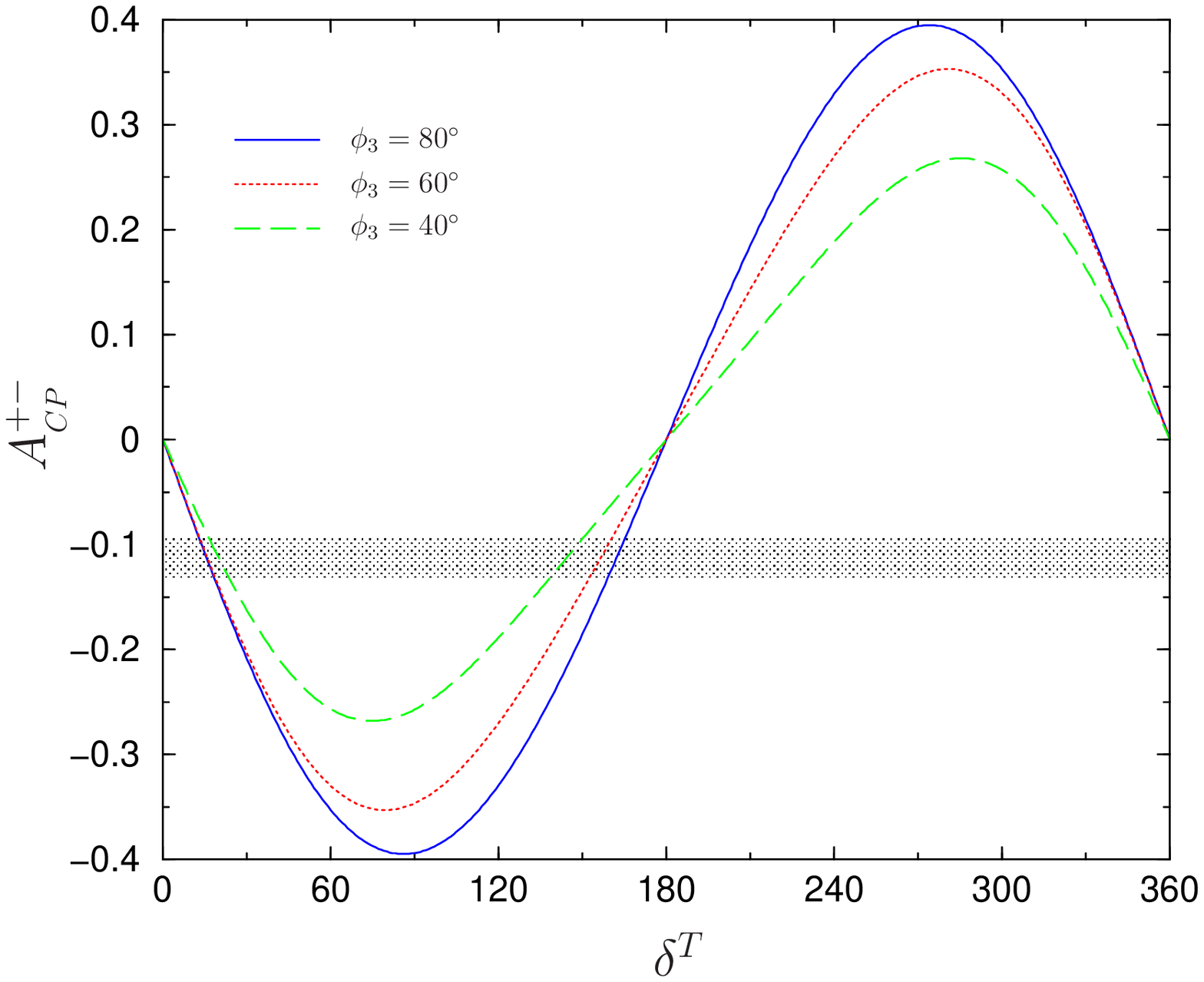}
 \caption{ $R_1$ and $A_{CP}^{+-}$ as a function of $\delta^T$ for three different
 values of $\phi_3 = 40^\circ, 60^\circ, 80^\circ$, respectively.
 The shaded regions denote the experimental limits.}
\label{fig:2}
\end{figure}

In Fig.~\ref{fig:2}, we present $R_1$ and $A_{CP}^{+-}$ as a function of $\delta^T$.  We
remind that both $R_1$ and $A_{CP}^{+-}$ depend only on $r_{_T}$, $\delta^T$ and $\phi_3$.
The left one of the figure shows $R_1$ as a function of $\delta^T$ for $\phi_3 = 40^\circ,
60^\circ, 80^\circ$, respectively, where $r_{_T}$ is fixed as $0.2$.  The allowed regions
are $\delta^T \leq 40^\circ$ or $\delta^T \geq 320^\circ$.  The possibility of the vanishing
$\delta^T$ is not excluded in this case.
In the right one of the figure, $A_{CP}^{+-}$ versus $\delta^T$ is presented for
$r_{_T} =0.2$ and $\phi_3 = 40^\circ, 60^\circ, 80^\circ$, respectively.  Combining the
results from the left and right ones, we find that the possibility of a large $\delta^T$ is
ruled out and the favored value of $\delta^T$ is non-zero and in between $20^\circ$ and
$30^\circ$ for $r_{_T} =0.2$ and $40^\circ \leq \phi_3 \leq 80^\circ$
\cite{Mishima:2004um,Yoshikawa:2003hb}.

\section{The $\chi^2$ analysis using $B \to K\pi$ decays}

In the numerical analysis, we use 8 observables, as shown in
Table I (except ${\cal A}_{CP}^{0+}$ which becomes irrelevant if
$r_{_A}$ is neglected).

{\bf Case (a):} In this case, we have eight observables as above and
seven parameters $( |\tilde P|,~ r_{_T},~ r_{_{EW}},~ r_{_C},~
\delta^T,~ \delta^{EW},~ \delta^C )$ so that the degrees of
freedom ({\it d.o.f.}) for the fit is 1.
The parameter values for the best fit are presented in Table~\ref{table:2}
for three different values of $\phi_3$ chosen from the unitarity triangle
fit: $\phi_3 = 54^\circ, ~ 60^\circ$, or $67^\circ$.
The minimum values of $\chi^2$ $(\chi^2_{min})$ in each case are also
shown in the table.

\begin{table}
\caption{ The theoretical parameter values for the best fit in {\bf Case (a)}. }
\smallskip
\begin{tabular}{|c|c|c|c|c|c|c|c|c|}
\hline
~$\phi_3$~~ &~~$|P|$ ~(in $10^{-6}$ GeV)~~&~~ $r_{_T}$~~
&~~ $r_{_{EW}}$~~ &~~$ r_{_C} $~~ &~~ $\delta^T $~~ &~~ $\delta^{EW}$~~
&~~ $\delta^C$ ~~ &~~ $\chi^2_{\rm min} / d.o.f.$ \\
\hline
${\it 54^\circ}$ & $1.23$ & $0.20$ & $0.35$ & $0.25$
& $17.9^\circ$ & $254.9^\circ$ & $195.1^\circ$ & $0.01$ \\
${\it 60^\circ}$ & $1.23$ & $0.25$ & $0.37$ & $0.25$
& $13.0^\circ$ & $260.4^\circ$ & $197.8^\circ$ & $0.005$ \\
${\it 67^\circ}$ & $1.22$ & $0.34$ & $0.40$ & $0.27$
& $8.6^\circ$ & $256.6^\circ$ & $202.2^\circ$ & $0.45$ \\
\hline
\end{tabular}
\label{table:2}
\end{table}

For instance, in the case of $\phi_3 = 60^\circ$, we find the best
fit with $\chi^2_{min}/ d.o.f. = 0.005$.  The corresponding parameters
are
\begin{eqnarray}
&& |\tilde P| = 1.23 \times 10^{-6} ~ {\rm GeV}, ~~ r_{_T} = 0.25, ~~
r_{_{EW}} = 0.37, ~~ r_{_C} = 0.25,  \\
&& \delta^T = 13.0^\circ, ~~ \delta^{EW} = 260.4^\circ, ~~
\delta^C = 197.8^\circ .
\end{eqnarray}
Using these parameter values, the observables are predicted as
\begin{eqnarray}
&& \bar {\cal B}^{0+} = 24.08 \times 10^{-6}, ~~~~
 \bar {\cal B}^{+0} = 12.10 \times 10^{-6},  \\
&& \bar {\cal B}^{+-} = 18.21 \times 10^{-6}, ~~~~
 \bar {\cal B}^{00} = 11.51 \times 10^{-6},  \\
&& {\cal A}_{CP}^{+0} = +0.04, ~~ {\cal A}_{CP}^{+-} = -0.119, ~~
 {\cal A}_{CP}^{00} = -0.007, ~~ S_{K_s \pi^0} = +0.33 ,
\end{eqnarray}
which are in good agreement with the central values of all the data.

\begin{figure}
 \centerline{ \DESepsf(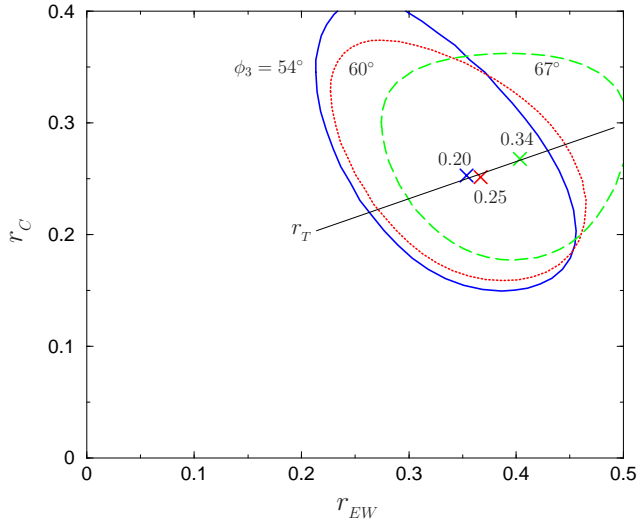 width 8cm)}
 \caption{{\bf Case (a)}: the allowed values of $r_{_{EW}}$ and $r_{_C}$ by
 the current data for $B \to K \pi$.
 The ``x'' marks denote the best fit values together with the corresponding values of
 $r_{_T}$ for $\phi_3 =54^\circ, ~60^\circ, ~67^\circ$, respectively.
 The regions surrounded by the solid, the dotted, and the dashed lines
 represent the allowed values of $r_{_{EW}}$ and $r_{_C}$ at 1$\sigma$
 level for $\phi_3 =54^\circ, ~60^\circ, ~67^\circ$, respectively.}
\label{fig:3}
\end{figure}

We see that in this case the best fit value of the color-suppressed tree contribution
$r_{_C}$ is comparable to that of the color-allowed tree contribution $r_{_T}$, and
the best fit value of the EW penguin contribution $r_{_{EW}}$ is also about 2.6 times
larger than the conventionally estimated one.
Table~\ref{table:2} shows that as $\phi_3$ increase from $54^\circ$ to $67^\circ$,
the best fit values of $r_{_T}$, $r_{_{EW}}$ and $r_{_C}$ also increases from 0.20, 0.35
and 0.25 to 0.34, 0.40 and 0.27, respectively.
The best fit indicates that in comparison to the conventional estimates within the SM,
the color-suppressed tree contribution should be enhanced by more than an order of
magnitude, and the EW penguin contribution needs to be enhanced up to a factor of 3.

In Fig.~\ref{fig:3}, the allowed values of $r_{_{EW}}$ and $r_{_C}$ are presented.
The ``x'' marks denote the best fit values together with the corresponding values of
$r_{_T} = 0.20, ~0.25, ~0.34$ for $\phi_3 =54^\circ, ~60^\circ, ~67^\circ$,
respectively.  The regions surrounded by the solid, the dotted, and the dashed lines
represent the allowed values of $r_{_{EW}}$ and $r_{_C}$ at 1$\sigma$ level for
$\phi_3 =54^\circ, ~60^\circ, ~67^\circ$, respectively.
It is obvious that the smallest allowed value of $r_{_C}$ at 1$\sigma$ level is at least
7 times larger than the conventional value ($\approx 0.02$), and the allowed value of
$r_{_{EW}}$ is also larger than its conventional SM estimate ($\approx 0.14$):
$r_{_{EW}} > 0.21$ for $54^\circ \leq \phi_3 \leq ~67^\circ$.
Therefore, the current experimental results for $B \to K\pi$ decays strongly imply
(large) enhancements of both the EW penguin and the color-suppressed tree contributions.

\begin{figure}
 \centerline{ \DESepsf(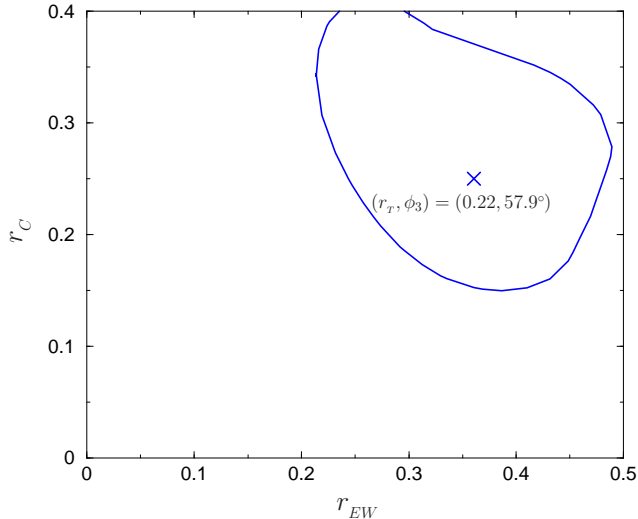 width 8cm)}
 \caption{{\bf Case (b)}: the allowed values of $r_{_{EW}}$ and $r_{_C}$ by the current
 data for $B \to K \pi$.  Here all the eight parameters, including $\phi_3$, are varied.
 The ``x'' mark denotes the best fit value where $r_{_T} =0.31$ and
 $\phi_3 =63^\circ$.}
\label{fig:4}
\end{figure}

{\bf Case (b):} Now we use all the eight parameters including $\phi_3$ together with the same
eight observables as before.  In this case, the $\phi_3$ varies from 54$^\circ$ to
67$^\circ$, which are chosen from the unitarity triangle fit.
The result at 1$\sigma$ level is shown in Fig.~\ref{fig:4}.
We note that this result confirms that of the case {\bf (iv)} (the dot-dashed line) shown
in Fig.~\ref{fig:1}.
The best fit values are
\begin{eqnarray}
&& |\tilde P| = 1.23 \times 10^{-6} ~ {\rm GeV}, ~~ r_{_T} = 0.22, ~~
r_{_{EW}} = 0.36, ~~ r_{_C} = 0.25,  \\
&& \delta^T = 14.8^\circ, ~~ \delta^{EW} = 258.5^\circ, ~~ \delta^C = 196.7^\circ,
~~ \phi_3 = 57.9^\circ .
\end{eqnarray}
The conclusion claimed in the case {\bf (a)} holds in this case as well.
That is, the present experimental results for $B \to K\pi$ decays strongly indicate simultaneous
(large) enhancements of the EW penguin and the color-suppressed tree contributions.

\begin{figure}[tb]
   \includegraphics[scale=0.45]{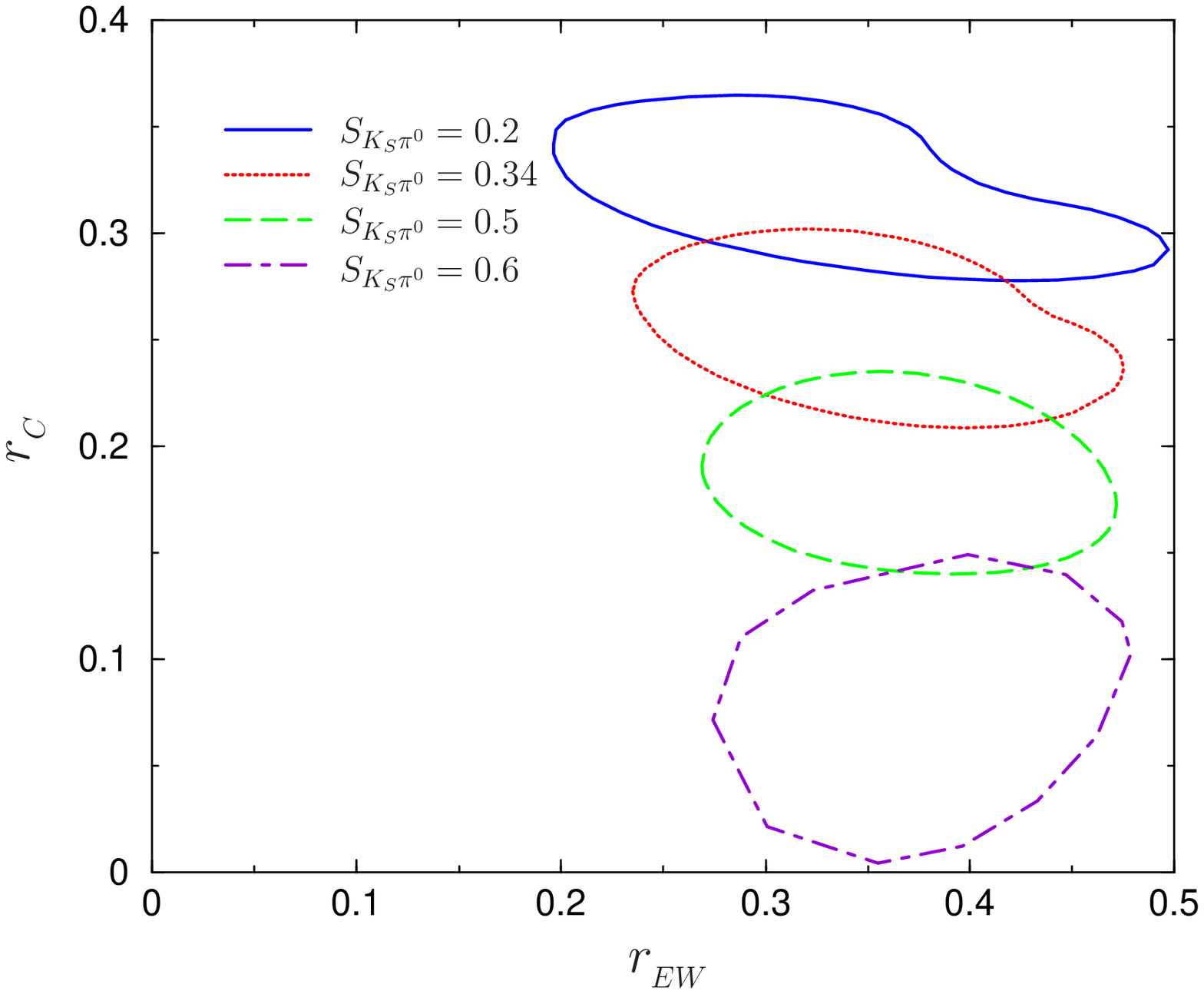}
 \hspace{1cm}
   \includegraphics[scale=0.45]{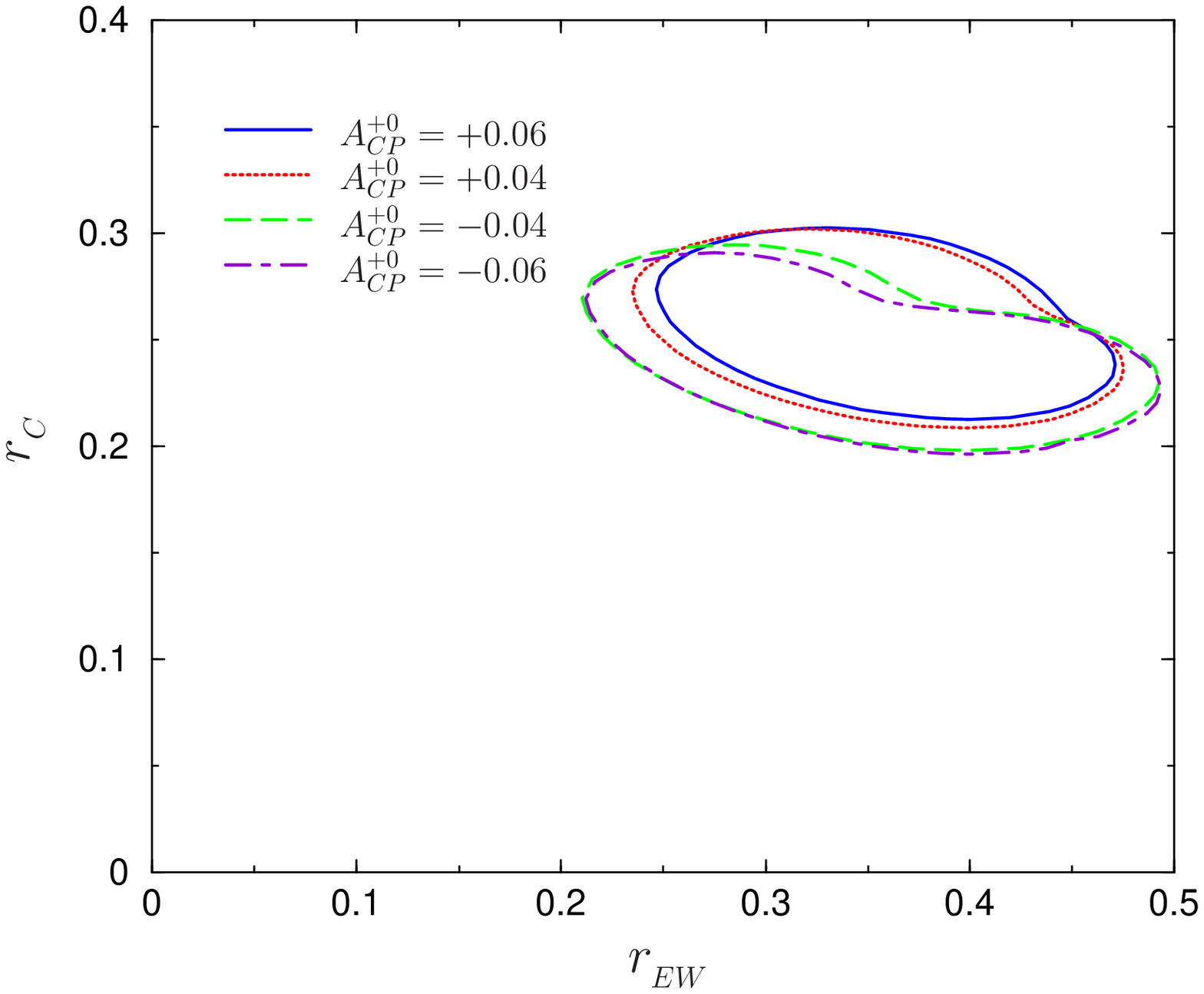}
 \caption{ Illustration of the allowed values of $r_{_{EW}}$ and $r_{_C}$ at 1$\sigma$ level
 by the current data for $B \to K \pi$, for a given $S_{K_s \pi^0}$ (left figure) or $A_{CP}^{+0}$
 (right figure).
 In the left figure, $S_{K_s \pi^0}$ is assumed to be $0.20, ~0.34, ~0.50, ~0.60$, respectively
 (plus 20\% error in each case).
 In the right one, $A_{CP}^{+0}$ is assumed to be $\pm 0.040, ~\pm 0.060$, respectively
 (plus 20\% error in each case).
 }
\label{fig:5}
\end{figure}

Finally we would like to make a comment on sensitivity between the parameter $r_{_C}$ and
the observable $S_{K_s \pi^0}$.
As implied in Eq. (\ref{Skpi}), the theoretical prediction of $S_{K_s \pi^0}$ can be sensitive to
the parameter $r_{_C}$.  For illustration, in the left one of Fig.~\ref{fig:5}, we show the allowed
values of $r_{_{EW}}$ and $r_{_C}$ at 1$\sigma$ level by the current data for $B \to K \pi$, when
the value of $S_{K_s \pi^0}$ changes, keeping the values of the other observables fixed
as in Table~\ref{table:1}.
We first vary $S_{K_s \pi^0}$ around the present experimental value:
specifically, $S_{K_s \pi^0}$ is assumed to be $(0.20 \pm 0.04)$, $(0.34 \pm 0.068)$,
$(0.50 \pm 0.10)$, $(0.60 \pm 0.12)$, respectively.
Here just for the illustrative purpose, we set 20\% errors in each case.
(Also, to be consistent, we set 20\% errors to all the data whose current errors are larger than
20\%, such as $A_{CP}^{ij}$.)
It is clear that as $S_{K_s \pi^0}$ varies, the allowed region for $r_{_C}$ varies sensitively:
as $S_{K_s \pi^0}$ increases, the allowed value of $r_{_C}$ decreases.
In contrast, $r_{_{EW}}$ is not sensitive to the change of $S_{K_s \pi^0}$.
Just for comparison, in the right one of Fig.~\ref{fig:5}, we also present the case that the value
of $A_{CP}^{+0}$ changes, keeping the values of the other ones fixed as in Table~\ref{table:1}.
Again for the illustrative purpose, $A_{CP}^{+0}$ is assumed to be $(+0.040 (-0.040) \pm 0.008)$,
$(0.060 (-0.060) \pm 0.012)$, respectively.
(To be consistent, we also set 20\% errors to all the data whose current errors are larger than
20\%, such as $S_{K_s \pi^0}$ and $A_{CP}^{ij}$.)
In contrast to the case of the left figure, both $r_{_{EW}}$ and $r_{_C}$ are insensitive to
the change of $A_{CP}^{+0}$.

\section{Conclusion}

We have studied the decay processes $B \to K\pi$ in a phenomenological way.
Using the currently available experimental data for all the $B \to K\pi$ modes, we have
determined the allowed values of the relevant theoretical parameters, such as
$|\tilde P|,~ r_{_T},~ r_{_{EW}},~ r_{_C},~ \delta^T,~ \delta^{EW},~ \delta^C, ~\phi_3$.
In order to find the most likely values of the parameters in a statically reliable way,
we used the $\chi^2$ analysis.

Our result shows that the current data for $B \to K\pi$ decays strongly indicate
(large) enhancements of both the EW penguin and the color-suppressed tree contributions:
$e.g.$, roughly, $0.21 \leq r_{_{EW}} \leq 0.49$ and $0.15 \leq r_{_C} \leq 0.43$
at 1$\sigma$ level.
The best fit values are $r_{_{EW}} = 0.36$ and $r_{_C} = 0.25$.
The favored values of $r_{_{EW}}$ and $r_{_C}$ are larger than the SM estimates
($r_{_{EW}} \approx 0.14$ and $r_{_C} \approx 0.02$) by about a factor of 2.5 and 12,
respectively.

It should be noted that in the case of using only the BRs ($i.e.$, not including CP asymmetries),
the conventional values of $r_{_{EW}}$ and $r_{_C}$ may not be completely excluded, if
the large value of $r_{_T}$ ($e.g.$, $r_{_T} \approx 0.4$) is assumed.

\vspace{1cm}
\centerline{\bf ACKNOWLEDGEMENTS}
\noindent
The work of C.S.K. was supported by in part by the Korea Research Foundation Grant funded
by the Korean Government (MOEHRD) No. R02-2003-000-10050-0.
The work of S.O. was supported by Korea Research Foundation Grant (KRF-2004-050-C00005).
The work of C.Y. was supported in part by Brain Korea 21 Program and
in part by Grant No. F01-2004-000-10292-0 of KOSEF-NSFC International
Collaborative Research Grant and in part by the Korea Research Foundation Grant funded
by the Korean Government (MOEHRD) No. R02-2003-000-10050-0.
\\


\end{document}